\documentclass[12pt]{article}
\usepackage{amsmath,amssymb,bm,epsfig}

\renewcommand{\thefootnote}{\fnsymbol{footnote}}

\begin{document}

\title{
\begin{flushright}
\begin{minipage}{0.2\linewidth}
\normalsize
WU-HEP-14-08 \\
EPHOU-14-017 \\*[50pt]
\end{minipage}
\end{flushright}
{\Large \bf 
Towards natural inflation \\from weakly coupled 
heterotic string theory 
\\*[20pt]}}

\author{Hiroyuki~Abe$^{1,}$\footnote{
E-mail address: abe@waseda.jp},\ \ 
Tatsuo~Kobayashi$^{2}$\footnote{
E-mail address:  kobayashi@particle.sci.hokudai.ac.jp}, \ and \ 
Hajime~Otsuka$^{1,}$\footnote{
E-mail address: hajime.13.gologo@akane.waseda.jp
}\\*[20pt]
$^1${\it \normalsize 
Department of Physics, Waseda University, 
Tokyo 169-8555, Japan} \\
$^2${\it \normalsize 
Department of Physics, Hokkaido University, Sapporo 060-0810, Japan} \\*[50pt]}

\date{
\centerline{\small \bf Abstract}
\begin{minipage}{0.9\linewidth}
\medskip 
\medskip 
\small
We propose the natural inflation from the 
heterotic string theory on ``Swiss-Cheese" 
Calabi-Yau manifold with multiple $U(1)$ magnetic 
fluxes. Such multiple $U(1)$ magnetic fluxes stabilize 
the same number of the linear combination of the 
universal axion and K\"ahler axions and one of the K\"ahler 
axions is identified as the inflaton. This axion decay constant 
can be determined by the size of one-loop corrections to the 
gauge kinetic function of the hidden gauge groups, 
which leads effectively to the trans-Planckian axion decay constant 
consistent with the WMAP, Planck and/or BICEP2 data. 
During the inflation, the real parts of the moduli 
are also stabilized by employing the nature of the 
``Swiss-Cheese" Calabi-Yau manifold. 
\end{minipage}
}

\begin{titlepage}
\maketitle
\thispagestyle{empty}
\clearpage
\tableofcontents
\thispagestyle{empty}
\end{titlepage}

\renewcommand{\thefootnote}{\arabic{footnote}}
\setcounter{footnote}{0}
\vspace{35pt}

\section{Introduction}
The cosmological inflation is an accelerated expansion 
of the universe at the early universe which can solve the 
horizon problem and the flatness problem at the same time. 
Such expanding universe is realized by the vacuum energy 
density of the scalar field, so called inflaton, whose 
quantum fluctuations produce the origin of the density 
perturbation of the universe. 

The current cosmological observations, especially, 
the Planck satellite report that the inflation scenario is 
well consistent with its data and the primordial density 
fluctuations are almost Gaussian and the spectrum index of the 
scalar density perturbation is nearly scale-invariant~\cite{Ade:2013uln}. 
Recently, the BICEP2 collaboration reported that the 
primordial tensor modes can be measured as B-mode 
polarization of the Cosmic Microwave Background (CMB), 
which leads to the tensor-to-scalar ratio, $r = {\cal O}(0.1)$, 
although there is a tension between the Planck and BICEP2 
collaborations.
 To achieve the tensor-to-scalar-ratio, $r = {\cal O}(0.1)$, 
we require that inflaton will roll slowly down to the minimum 
of its scalar potential from its Planckian field value which is 
problematic from the theoretical point of view, especially in 
the string theory which will be expected as the unified theory 
of gauge and gravitational interactions. 

In the higher dimensional theories as well as the string theories, 
there are a lot of axions associated with the internal cycles of 
the internal manifolds such as the Calabi-Yau (CY) manifold which 
keeps the only ${\cal N}=1$ supersymmetry (SUSY) in the 
four-dimensional ($4$D) spacetime. 
When we consider such axions as the candidate of inflaton, 
the natural inflation~\cite{Freese:1990rb} is an attractive 
scenario as one of the large-field inflation, which is 
originally proposed by identifying the inflaton as the pseudo-Nambu 
Goldstone boson \footnote{The axion monodromy inflation is another interesting possibility.
 See e.g. \cite{Silverstein:2008sg,Kaloper:2008fb,Kobayashi:2014ooa}.}. 
However, the natural inflation compatible with the observed 
Planck~\cite{Ade:2013uln} and/or BICEP2~\cite{Ade:2014xna} 
data requires the trans-Planckian axion decay constant, 
see Ref.~\cite{Freese:2014nla} and references therein. 
So far, there are several approaches to realize the natural 
inflation with trans-Planckian axion decay constant 
in the framework of supergravity models or 
Type IIB superstring theory~\cite{Kallosh:2007ig,Czerny:2014xja,Choi:2014rja,Tye:2014tja,Kappl:2014lra,Ben-Dayan:2014zsa,Long:2014dta,Abe:2014vca,Li:2014lpa}, 
although, in the string theory, the fundamental axion decay 
constants are typically in the range 
$10^{16}- 10^{17} {\rm GeV}$~\cite{Choi:1985je}.  

In this paper, we propose the single-field natural inflation in the framework 
of heterotic string theory on ``Swiss-Cheese" CY 
manifold with multiple $U(1)$ fluxes induced from the anomalous 
$U(1)$ symmetries. 
By employing such multiple $U(1)$ fluxes, the linear combination of 
the universal axion and K\"ahler axions except for the inflaton are 
absorbed by the multiple $U(1)$ gauge bosons and they get the mass terms 
from these $U(1)$ fluxes. From the phenomenological point of view, 
$U(1)$ fluxes may be important tools to realize the $4$D standard 
model gauge groups from the heterotic string 
theory~\cite{Witten:1984dg,Blumenhagen:2006ux} 
as well as the Type IIB superstring theory~\cite{Cremades:2004wa}. 
During and after the inflation, the dilaton and real parts of K\"ahler 
moduli have to be stabilized and decoupled from the inflaton 
dynamics, otherwise 
the oscillations of these moduli would lead to the sizable isocurvature 
perturbations. In our model, the stabilization of the dilaton and real 
parts of K\"ahler moduli are realized by the non-perturbative corrections to 
the K\"ahler potential, superpotential and a nature of the structure of 
 ``Swiss-Cheese" CY manifold whose manifolds 
are also well studied as several topics about the particle phenomenology 
and cosmology based on the Type IIB string theory~\cite{Balasubramanian:2005zx} 
or F-theory~\cite{Blumenhagen:2009gk} and moduli stabilization based 
on the heterotic string theory~\cite{Cicoli:2013rwa}. 

This paper is organized as follows. In Sec.~\ref{sec:hetero}, 
we review the heterotic string theory on CY manifold with 
multiple $U(1)$ magnetic fluxes. 
We propose two inflation models in Secs.~\ref{subsec:naturalinf1} 
and \ref{subsec:naturalinf2}. 
Both are consistent with the WMAP, Planck and/or BICEP2 
data. Sec.~\ref{sec:con} is devoted to the conclusion. 
In Appendix~\ref{app:mass}, we show the mass matrices of fields 
for inflation model $1$ in Sec.~\ref{subsec:naturalinf1}.

\section{Heterotic string on CY manifolds with multiple 
$U(1)$ magnetic fluxes}
\label{sec:hetero}
We consider the $E_8\times E_8$ or $SO(32)$ heterotic string theory 
on Calabi-Yau manifold with multiple $U(1)$ magnetic fluxes (in other 
words, multiple line bundles). The low-energy effective theory of the 
heterotic string is given by the following Lagrangian at the string 
frame,
\begin{align}
S_{\rm bos}&=\frac{1}{2\kappa_{10}^2}\int_{M^{(10)}} 
e^{-2\phi_{10}} \left[ R+4d\phi_{10} \wedge  \ast 
d\phi_{10} -\frac{1}{2}H\wedge \ast H \right] 
\nonumber\\
&-\frac{1}{2g_{10}^2} \int_{M^{(10)}} e^{-2\phi_{10}} 
{\rm tr}(F\wedge \ast F),
\label{eq:heterob}
\end{align}
which is the bosonic part of the Lagrangian in the notation 
of~\cite{Polchinsky} and $\phi_{10}$ is the dilaton and $F$ 
is the field strength of $E_8\times E_8$ or $SO(32)$ 
gauge groups and then ``tr" denotes in the vector 
representation of these gauge groups. As will be mentioned later, 
the $U(1)$ magnetic fluxes are inserted in these 
gauge groups. $H$ is the three-form field strength defined 
by 
\begin{align}
H&=dB^{(2)} -\frac{\alpha^{'}}{4}(w_{\rm YM} -w_{L}),
\end{align}
where $w_{\rm YM}$ and $w_{L}$ are the gauge and 
gravitational Chern-Simons three-form, respectively.
The gravitational and Yang-Mills couplings are 
set by $2\kappa_{10}^2 =(2\pi)^7 (\alpha^{'})^4$ and 
$g_{10}^2 =2(2\pi)^7 (\alpha^{'})^3$. 

Throughout this work, we focus on the 
$E_8^{\rm vis}\times E_8^{\rm hid}$ heterotic string with 
non-standard embedding, that is, the visible $E_8^{\rm vis}$ 
gauge group decomposes into the product group of 
$G_{\rm vis}$ and multiple $U(1)$s where $G_{\rm vis}$ 
is the Grand Unified Group (GUT) or just the stand model (SM)
gauge groups and we do not consider the charged scalar 
fields under the multiple $U(1)$s 
\footnote{An extension to the cases for $SO(32)$ 
heterotic string theory is straightforward.}. 
In addition to it, we assume that the hidden gauge groups 
are just non-abelian gauge groups, for simplicity.

After the dimensional reduction on the CY manifold with 
multiple $U(1)$ magnetic fluxes, 
we get the following $4$D $U(1)$ 
invariant effective tree-level K\"ahler potential,
\begin{align}
{\cal K}=&-M_{\rm Pl}^2
\Biggl[
\ln \left( S+\bar{S} -\sum_{m} \frac{{\cal Q}_S^m}{16\pi^2}V_m \right) 
\nonumber\\
&+\ln \Biggl\{\frac{d_{ijk}}{48} 
\left( T_i +\bar{T}_i -\sum_{m} \frac{{\cal Q}_{T_i}^m}{2\pi} V_m \right)
\left( T_j +\bar{T}_j -\sum_{m} \frac{{\cal Q}_{T_j}^m}{2\pi} V_m \right)  
\left( T_k +\bar{T}_k -\sum_{m} \frac{{\cal Q}_{T_k}^m}{2\pi} V_m \right) 
\Biggl\}\Biggl],
\label{eq:Kahler}
\end{align} 
where $M_{\rm Pl}^2=\frac{e^{-2\phi_{10}}{\cal V}}{\kappa_{10}^2}$, 
$m$ labels the number of anomalous $U(1)$ vector multiplets 
$V_m$, $d_{ijk}$ are the intersection 
numbers of the Calabi-Yau manifold.
$S$ and $T_i$ for $i=1,2,\cdots h^{1,1}$ are the superfield 
descriptions of the dilaton and the K\"ahler moduli, respectively,
\begin{align}
S&=\frac{1}{4\pi}\left[ \frac{e^{-2\phi_{10}}{\cal V}}{l_s^6} 
+i b_S^{(0)}\right],\nonumber\\\
T_i&=t_i+i b_{T_i}^{(0)},
\end{align} 
where ${\cal V}=\frac{1}{6}\int_{\rm CY}J\wedge J\wedge J$ with 
$J=l_s^2\sum_{i}t_iw_i$ is the volume of the 
CY manifold, $J$ is the K\"ahler form expanded by the base of 
two-form $w_i$, $i=1,\cdots, h^{1,1}$ and $l_s=2\pi\sqrt{\alpha^\prime}$ is the string length. 
The imaginary parts of $S$ and $T_i$, $b_S^{(0)}$ and 
$b_{T_i}^{(0)}$ are the universal and K\"ahler axions given by the 
dimensional reduction of the Kalb-Ramond two-form $B^{(2)}$ 
and six-form $B^{(6)}$ as 
\begin{align}
&B^{(2)} =b_S^{(2)} +l_s^2\sum_{k=1}^{h_{11}} b_{T_k}^{(0)} w_k,
\nonumber\\
&B^{(6)} =l_s^6b_S^{(0)}{\rm vol}_6 +l_s^4\sum_{k=1}^{h_{11}} 
b_{T_k}^{(2)} \hat{w}_k, 
\end{align} 
where $b_S^{(2)}$ and $b_{T_i}^{(2)}$ are the $4$D tensor fields, 
${\rm vol}_6$ is the normalized volume form, 
$\int_{\rm CY}{\rm vol}_6=1$, and 
$\hat{w}_i$ are the Hodge dual four-form of the two-form $w_i$. 
The two-form $B^{(2)}$ and six-form $B^{(6)}$ are related 
by the Hodge duality, $\ast_{10} dB^{(2)}=e^{2\phi_{10}}dB^{(6)}$ 
and $\ast_{4}db_I^{(2)}=e^{2\phi_{10}}db_I^{(0)}$, 
$I=S,T_1,\cdots T_{h^{1,1}}$. 

The $U(1)$ charges of the dilaton and K\"ahler moduli for 
the $U(1)^m$ symmetries, ${\cal Q}_I^m$, $I=S,T_1,\cdots T_{h^{1,1}}$ 
are defined via 
the following couplings of the $U(1)$ gauge bosons $A_m$,
\begin{align}
S \supset 
\sum_{m}\frac{{\cal Q}_S^m}{4l_s^2}
\int_{R^{1,3}}b_S^{(2)} \wedge F_m +
\sum_{i,m}\frac{{\cal Q}_{T_i}^m}{2l_s^2}
\int_{R^{1,3}}b_{T_i}^{(2)} \wedge F_m,
\label{eq:masscp}
\end{align} 
where 
\begin{align}
{\cal Q}_S^m\equiv {\rm tr}(T^mT^m)\int_{\rm CY}\frac{{\rm tr}\bar{F}_m}{2\pi} 
\wedge \frac{1}{16\pi^2} 
\left( {\rm tr}\bar{F}^2 -\frac{1}{2}{\rm tr}\bar{R}^2\right),
\,\,\,
{\cal Q}_{T_i}^m\equiv {\rm tr}(T^mT^m)\int_{T_i}\frac{{\rm tr}\bar{F}_m}{2\pi}, 
\label{eq:charge}
\end{align} 
$T^m$ are the $U(1)^m$ generators embedded in the visible $E_8$ 
gauge group, ${\bar F}_m$ and ${\bar F}$ are the internal field 
strengths of the $U(1)^m$ symmetry and $E_8^{\rm vis}$ symmetry. 
Such couplings are obtained from the dimensional reduction 
of the $10$D kinetic terms of $H$ given by Eq.~(\ref{eq:heterob}) 
and the one-loop Green-Schwarz (GS) 
counter term~\cite{Ibanez:1986xy} 
which is determined by the S-dual of the 
type I theory as shown in Appendix of Ref.~\cite{Blumenhagen:2006ux}, 
\begin{align}
S_{\rm GS}=\frac{1}{24(2\pi)^5\alpha'}\int B^{(2)}\wedge X_8,
\label{eq:GS}
\end{align}
where the eight-form $X_8$ reads,
\begin{align}
X_8=\frac{1}{24}{\rm Tr}F^4-\frac{1}{7200}({\rm Tr}F^2)^2-
\frac{1}{240}({\rm Tr}F^2)({\rm tr}R^2) +\frac{1}{8}{\rm tr}R^4 
+\frac{1}{32}({\rm tr}R^2)^2.
\end{align}

The mass terms of the $U(1)$ gauge 
bosons are derived by expanding the K\"ahler 
potential to second order on the vector multiplets,
\begin{align}
S_{\rm mass}&=-\sum_{m,n} \frac{M_{\rm Pl}^2}{4}
\left( \frac{K_{S\bar{S}}{\cal Q}_S^m {\cal Q}_S^n}{(16\pi^2)^2}
+\sum_{i,j} \frac{K_{T_i\bar{T_j}}{\cal Q}_{T_i}^m {\cal Q}_{T_j}^n}{(2\pi)^2}\right)
\int_{R^{1,3}} A_{m} \wedge \ast_4 A_n,
\label{eq:mass}
\end{align}
which is typically of order the 
string scale $M_s^2=1/l_s^2$ with $l_s=2\pi\sqrt{\alpha^{\prime}}$, 
see Refs.~\cite{Blumenhagen:2005ga} for $E_8 \times E_8$ and~\cite{Blumenhagen:2005pm} for $SO(32)$ 
heterotic string theories and references therein.

From the $U(1)$ invariant K\"ahler potential given by 
Eq.~(\ref{eq:Kahler}), $U(1)$ magnetic fluxes generate 
the moduli-dependent Fayet-Iliopoulos terms~\cite{Atick:1987gy}, 
\begin{align}
\xi_m &=
\frac{\partial{\cal K}}{\partial V_m}\Biggl|_{V_m=0} 
=-\frac{{\cal Q}_S^m}{16\pi^2} K_S -\sum_{i=1}^{h^{1,1}} 
\frac{{\cal Q}_{T_i}^m}{2\pi} K_{T_i},
\end{align} 
where $K_I=\partial K/\partial Z^I$ for $Z^I=S,T_1,\cdots, T_{h^{1,1}}$. 
Finally, we comment on the gauge kinetic function of the 
non-abelian gauge groups obtained from the decomposition 
of the $E_8^{({\rm vis})}\times E_8^{({\rm hid})}$ heterotic 
string theory. They receive the one-loop 
corrections originating from the one-loop GS 
term as shown in Eq.~(\ref{eq:GS}),
\begin{align}
&f_{\rm vis} =S+\beta_i T_i, \nonumber\\
&f_{\rm hid} =S-\beta_i T_i,
\label{eq:gaugekin}
\end{align} 
where 
\begin{align}
\beta_i \equiv \frac{1}{8\pi}\int_{\rm CY}
\frac{1}{16\pi^2} 
\left( {\rm tr}\bar{F}^2 -\frac{1}{2}{\rm tr}\bar{R}^2\right)
\wedge \hat{w}_i.
\label{eq:oneloop}
\end{align} 
Both gauge kinetic functions in the visible and 
hidden sector are correlated by the tadpole 
cancellation condition of the 
$E_8^{({\rm vis})}\times E_8^{({\rm hid})}$ heterotic string theory. 
For the $SO(32)$ heterotic string theory, the 
non-abelian gauge groups included in $SO(32)$ have 
the nonuniversal gauge kinetic functions depend 
on the decomposition of $SO(32)$.

\section{Natural inflation from heterotic string theory}
\label{sec:naturalinf}
In this section, we propose two natural inflation scenarios 
in the framework of the weakly coupled heterotic string 
theory on ``Swiss-Cheese" Calabi-Yau manifold with multiple $U(1)$ 
fluxes induced from the anomalous $U(1)$ symmetries. 
As pointed out in the introduction, 
both natural inflation scenarios are the single-filed 
inflation models whose inflaton are identified as the single 
K\"ahler axion with trans-Planckian axion decay constant. 
The trans-Planckian axion decay constant is originating 
from the one-loop corrections to the gauge kinetic function 
of the hidden gauge groups to achieve the successful natural 
inflation which is different from the natural 
inflation scenarios by employing two axions with sub-Planckian 
axion decay constants~\cite{Kim:2004rp}. 

On the other hand, the other K\"ahler 
axions are absorbed by the multiple $U(1)$ gauge bosons 
and become massive. The real 
part of dilaton is stabilized at the finite value 
by the contributions from the non-perturbative effect 
to the dilaton K\"ahler potential and gaugino condensation 
term as shown in Secs.~\ref{subsec:naturalinf1} 
and \ref{subsec:naturalinf2}, respectively. 
One of the real parts of K\"ahler moduli is stabilized by 
the world sheet instanton effect which leads to the 
stabilization of other real parts of K\"ahler moduli. 
 
%The volume form of ``Swiss-Cheese" Calabi-Yau is expressed as 
%\begin{align}
%{\cal V}=t_b^3-\sum_{i=1}^{h^{1,1}-1}\gamma_i t_{s^i}^3,
%\end{align} 
%where $\gamma_i$ are positive constants. 
\subsection{Model $1$ (Single gaugino condensation)}
\label{subsec:naturalinf1}
In this section, we will show the inflaton potential along 
the following three steps. First, the universal axion 
and K\"ahler axions except for the inflaton 
are absorbed by the multiple $U(1)$ gauge bosons 
at the string scale as shown in Eq.~(\ref{eq:mass}). 
Next, the dilaton and all real parts of K\"ahler moduli 
are stabilized at the SUSY breaking minimum 
by the inclusion of non-perturbative corrections 
to the dilaton K\"ahler potential and superpotential 
which is the world sheet instanton effect. 
Finally, below the SUSY breaking scale, we get 
the effective scalar potential for the light K\"ahler 
axion which is identified as the inflaton later. 

\subsubsection{Setup}
\label{subsubsec:naturalinf1}
We consider the following K\"ahler potential with five 
K\"ahler moduli and four 
anomalous $U(1)$ symmetries, 
\begin{align}
{\cal K}=&K
\left( S+\bar{S}, V^1, V^2, V^3\right)
\nonumber\\
-&\ln \left\{ k_1(T_1+\bar{T}_1)^3 
-k_2\left( T_2 +\bar{T}_2 -\sum_{n=1}^3q_{T_2}^n V^n\right)^3 
-k_3\left( T_3 +\bar{T}_3 -\sum_{n=1}^3q_{T_3}^n V^n\right)^3 \right.
\nonumber\\
&\left. -k_4\left(T_4 +\bar{T}_4 -q_{T_4}^4 V^4 \right)^3
-k_5\left(T_5 +\bar{T}_5 -q_{T_5}^4 V^4 \right)^3 
\right\},
\label{eq:m1K}
\end{align} 
in the unit $M_{\rm Pl}=1$ 
with $M_{\rm Pl}=2.4\times 10^{18} {\rm GeV}$, 
where we choose $h^{1,1}=5$, $q_{T_i}^m={\cal Q}_{T_i}^m/2\pi$, 
$i=1,2,3,4,5$, $m=1,2,3,4$ and $k_i$, $i=1,2,3,4,5$ 
are the positive constants determined by the intersection 
numbers $d_{t_1t_1t_1}$, $d_{t_2t_2t_2}$, $d_{t_3t_3t_3}$, 
$d_{t_4t_4t_4}$,$d_{t_5t_5t_5}$. (The reason why we 
choose five K\"ahler moduli and four anomalous $U(1)$s are 
shown later.)

The K\"ahler potential of dilaton consists of the tree-level 
and the non-perturbative part,
\begin{align}
K^0&=-
\ln \left( S+\bar{S}-\sum_{n=1}^3q_S^n V^n\right),
\nonumber\\
K^{\rm np}&=
d\,g^{-p}
e^{-b/g},
\label{eq:dilaton}
\end{align} 
where $b$, $p$, and $d$ are the real constants, 
$q_s^n={\cal Q}_S^n/16\pi^2$, $n=1,2,3$ and 
$g=({\rm Re}\,S-\sum_{i\neq 1} \beta_i{\rm Re}\,T_i)^{-1/2}$ 
is the gauge coupling in the hidden sector as 
shown in the superpotential (\ref{eq:gaugekin}). 
$K^{\rm np}$ denotes the non-perturbative correction to the 
K\"ahler potential~\cite{Shenker:1990uf,Burgess:1995aa,Casas:1996zi}. 
There are known ansatz to write the dilaton K\"ahler 
potential as 
\begin{align}
K&=K^0+K^{\rm np}\,\,\,{\rm or}\,\,\,\,\,
K=\ln \left( e^{K^0}+e^{K^{\rm np}}\right),
\label{eq:dilatonK}
\end{align} 
etc.\footnote{In~\cite{Burgess:1995aa}, the non-perturbative 
K\"ahler potential of dilaton is discussed in the effective field 
theory approaches.}. Anyway, we assume that the dilaton is stabilized at 
the finite value due to such corrections to the K\"ahler 
potential as discussed in~\cite{Higaki:2003jt}.
Note that our following moduli stabilization as well as the 
inflation mechanism do not depend on the detailed structure 
of the non-perturbative K\"ahler potential, $K^{\rm np}$. 

In addition to the K\"ahler potential, we consider the 
following $U(1)^m$, $m=1,2,3,4$, invariant superpotential,
\begin{align}
W=&W_0+A\,e^{-\frac{8\pi^2}{a}(S-\beta_2 T_2-\beta_3 T_3
-\beta_4 T_4-\beta_5 T_5)} 
+B\,e^{-\mu_1 T_1}, 
\label{eq:m1W}
\end{align} 
where $W_0$ is the Neveu-Schwarz (NS) three-form flux induced constant 
term which stabilizes 
the $h^{1,2}$ complex structure moduli of the CY manifold, 
the second term of the right handed side (r.h.s.) shows the 
hidden sector gaugino condensation which 
receive the one-loop corrections originating from the one-loop 
Green-Schwarz counter term given by Eq.~(\ref{eq:oneloop}). 
The third term of the (r.h.s.) shows the world-sheet instanton 
effects on the two cycle $T_1$.  

As the first step to obtain the inflaton potential, 
we comment on the anomalous $U(1)^m$ 
vector multiplets $V^m$, $m=1,2,3,4$. 
Such $U(1)^m$ vector multiplets are massive due to the 
$U(1)^m$ magnetic fluxes as shown in Eq.~(\ref{eq:mass}) and then 
$U(1)^m$ gauge bosons absorb the linear 
combination of imaginary component of the dilaton and the K\"ahler 
moduli. 
$U(1)^1$, $U(1)^2$ and $U(1)^3$ gauge bosons absorb 
the following linear combination of the moduli, 
\begin{align}
X^1&=\frac{1}{N^1}\left( \frac{{\rm Im}\,S}{q_S^1\sqrt{K_{S\bar{S}}}} 
+\frac{{\rm Im}\,T_2}{q_{T_2}^1\sqrt{K_{T_2\bar{T}_2}}}
+\frac{{\rm Im}\,T_3}{q_{T_3}^1\sqrt{K_{T_3\bar{T}_3}}}\right), 
\nonumber\\
X^2&=\frac{1}{N^2}\left( \frac{{\rm Im}\,S}{q_S^2\sqrt{K_{S\bar{S}}}} 
+\frac{{\rm Im}\,T_2}{q_{T_2}^2\sqrt{K_{T_2\bar{T}_2}}}
+\frac{{\rm Im}\,T_3}{q_{T_3}^2\sqrt{K_{T_3\bar{T}_3}}}\right), 
\nonumber\\
X^3&=\frac{1}{N^3}\left( \frac{{\rm Im}\,S}{q_S^3\sqrt{K_{S\bar{S}}}} 
+\frac{{\rm Im}\,T_2}{q_{T_2}^3\sqrt{K_{T_2\bar{T}_2}}}
+\frac{{\rm Im}\,T_3}{q_{T_3}^3\sqrt{K_{T_3\bar{T}_3}}}\right), 
\end{align} 
where $N^i=\sqrt{(1/q_S^n\sqrt{K_{S\bar{S}}})^2+(1/q_{T_2}^n\sqrt{K_{T_2\bar{T}_2}})^2+(1/q_{T_3}^n\sqrt{K_{T_3\bar{T}_3}})^2}$, $n=1,2,3$,
respectively. Here the dilaton and K\"ahler moduli 
are canonically normalized and their K\"ahler metric 
are summarized in Appendix~\ref{app:mass}.

Thus, the imaginary component of $S$, $T_2$ and $T_3$ 
are absorbed by the $U(1)^1$, $U(1)^2$, $U(1)^3$ gauge bosons 
and their mass-squared matrices are given by
\begin{align}
M_{m,n}^2\simeq \frac{M_{\rm Pl}^2}{4\sqrt{\langle{\rm Re}\,f_{m,m}\rangle}
\sqrt{\langle{\rm Re}\,f_{n,n}\rangle}}
\left( K_{S\bar{S}}q_S^m q_S^n
+\sum_{i,j} K_{T_i\bar{T_j}}q_{T_i}^m q_{T_j}^n\right)
\label{eq:m1mass}
\end{align} 
for $m,n=1,2,3$, where the $U(1)$ gauge bosons are canonically 
normalized. The gauge kinetic functions 
of $U(1)$s, $f_{m,n}$ are given by 
$f_{m,n}={\rm tr}(T^mT^n)S\delta_{m,n} +{\cal O}(\beta T)$. 
These $U(1)^{1}$, $U(1)^{2}$ charges of the moduli 
$S$, $T_2$ and $T_3$ are related by the $U(1)^{1}$, $U(1)^{2}$ 
and $U(1)^{3}$ gauge invariance of the superpotential~(\ref{eq:m1W}), 
\begin{align}
q_S^1=q_{T_2}^1\,\beta_2 +q_{T_3}^1\,\beta_3,\,\,\,
q_S^2=q_{T_2}^2\,\beta_2 +q_{T_3}^2\,\beta_3,\,\,\,
q_S^3=q_{T_2}^3\,\beta_2 +q_{T_3}^3\,\beta_3,
\label{eq:cdcharge}
\end{align} 
Under the $U(1)$ gauge invariance condition~(\ref{eq:cdcharge}), 
the full-rank mass matrices~(\ref{eq:m1mass}) 
are realized if the number of $U(1)$s are bigger than three. 
Thus we can stabilize the K\"ahler axion and 
universal axion. The universal axion cannot be identified as
the candidate of inflaton, because its decay constant is 
much less than the Planck scale as shown in the 
superpotential~(\ref{eq:m1W}).  

The other $U(1)^4$ gauge boson absorbs the following 
combination of the moduli, 
\begin{align}
X^4=\frac{1}{N^4}\left( \frac{{\rm Im}\,T_4}{q_{T_4}^4\sqrt{K_{T_4\bar{T}_4}}} 
+\frac{{\rm Im}\,T_5}{q_{T_5}^4\sqrt{K_{T_5\bar{T}_5}}}\right), 
\end{align} 
where $N^4=\sqrt{(1/q_{T_4}^4\sqrt{K_{T_4\bar{T}_4}})^2+(1/q_{T_5}^4\sqrt{K_{T_5\bar{T}_5}})^2}$, and 
the orthogonal direction of $X_4$ (which is 
identified as the inflaton later), 
\begin{align}
Y^4=\frac{1}{N^4}\left(-\frac{{\rm Im}\,T_4}{q_{T_5}^4\sqrt{K_{T_5\bar{T}_5}}} 
+\frac{{\rm Im}\,T_5}{q_{T_4}^4\sqrt{K_{T_4\bar{T}_4}}} \right), 
\label{eq:infdef}
\end{align} 
cannot be absorbed by the anomalous $U(1)$ gauge 
bosons and the mass of $Y^4$ is obtained from the 
gaugino condensation term in Eq.~(\ref{eq:m1W}). 
In summary, 
the four imaginary parts of the moduli $X^m$, $m=1,2,3,4$ are 
absorbed by four anomalous $U(1)$ vector multiplets 
after following the above procedure. 
As the second step to obtain the inflaton potential, 
let us discuss the F-term potential derived from the 
K\"ahler potential~(\ref{eq:m1K}) 
and the superpotential~(\ref{eq:m1W}). 

First, we redefine the linear combination of 
the dilaton and the K\"ahler moduli as,
\begin{align}
\Phi =S-\beta_2 T_2-\beta_3 T_3-\beta_4 T_4-\beta_5 T_5,
\end{align} 
and then the K\"ahler potential and superpotential given by 
Eqs.~(\ref{eq:m1K}) and (\ref{eq:m1W}) are rewritten by 
\begin{align}
{\cal K}=&K\left( \Phi +\bar{\Phi}, T_2+\bar{T}_2, T_3 +\bar{T}_3, 
T_4+\bar{T}_4, T_5 +\bar{T}_5, V^1, V^2, V^3\right) \nonumber\\ 
-&\ln \left\{ k_1(T_1+\bar{T}_1)^3 
-k_2\left( T_2 +\bar{T}_2 -\sum_{n=1}^3q_{T_2}^n V^n\right)^3 
-k_3\left( T_3 +\bar{T}_3 -\sum_{n=1}^3q_{T_3}^n V^n\right)^3 \right.
\nonumber\\
&\left. -k_4\left(T_4 +\bar{T}_4 -q_{T_4}^4 V^4 \right)^3
-k_5\left(T_5 +\bar{T}_5 -q_{T_5}^4 V^4 \right)^3 
\right\},
\nonumber\\
W=&\,W_0+A\,e^{-\frac{8\pi^2}{a}\,\Phi} 
+B\,e^{-\mu_1 T_1}, 
\label{eq:m1KW}
\end{align} 
and we assume that the gaugino condensation term 
in Eq.~(\ref{eq:m1KW}) is much smaller than the other 
terms in Eq.~(\ref{eq:m1KW}) at least at the minimum, that is, 
$W_0$, $B\,e^{-\mu_1 T_1}$ 
$\gg Ae^{-\frac{8\pi^2}{a}\,\Phi}$. 
Therefore, at the moment, we ignore the contribution 
of the gaugino condensation term as we will mention later. 

Second, we stabilize the moduli $T_1$, ${\rm Re}\,T_2$, 
${\rm Re}\,T_3$, ${\rm Re}\,T_4$, ${\rm Re}\,T_5$ and ${\rm Re}\,\Phi$
by imposing the supersymmetric conditions,
\begin{align}
&D_{T_1}W=0,\nonumber\\
&D_{T_2}W=K_{T_2}W=0,\,\,
D_{T_3}W=K_{T_3}W=0, \,\,
D_{T_4}W=K_{T_4}W=0,\,\,
D_{T_5}W=K_{T_5}W=0 \nonumber\\
&D_{\Phi}W=K_{\Phi}W=0,
\label{eq:m1cd1}
\end{align} 
where $D_I W=W_I +K_I W$ and $W_I=\partial W/\partial Z_I$ 
with $Z_I=T_1,T_2,T_3,T_4,T_5,\Phi$. 
Then the D-term potential induced from the 
K\"ahler potential~(\ref{eq:m1K}) 
are automatically vanished under 
the above supersymmetric conditions, 
$K_{T_2}=K_{T_3}=K_{T_4}=K_{T_5}=K_{\Phi}=0$. 
${\rm Re}\,\Phi$ is stabilized by the contribution from the 
non-perturbative correction to the dilaton,
\begin{align}
K_{\Phi}=0.
\end{align}  
In the same way for ${\rm Re}\,\Phi$, the real parts of moduli 
$T_j$, $j=2,3,4,5$, are stabilized by the following conditions,
\begin{align}
K_{T_j}\simeq \frac{3k_j(T_j+\bar{T}_j)^2}{k_1(T_1+\bar{T}_1)^3}
+\frac{\partial K^0}{\partial T_j} 
\simeq \frac{3k_j(T_j+\bar{T}_j)^2}{k_1(T_1+\bar{T}_1)^3}
-\frac{\beta_j}{\Phi +{\bar \Phi}}
+{\cal O}\left(\beta_j\frac{\sum_{k=2}^5 \beta_{k}{\rm Re}\,T_k}{{\rm Re}\,\Phi}\right) 
=0,
\end{align}  
where the dilaton K\"ahler potential is approximated as 
its tree-level part $K^0$ in Eq.~(\ref{eq:dilatonK}). 
In the limit of ${\rm Re}\,T_1 >{\rm Re}\,T_j$, 
$j=2,3,4,5$, the above equations are rewritten by 
\begin{align}
{\rm Re}\,S\simeq {\rm Re}\,\Phi \simeq 
\frac{k_1({\rm Re}\,T_1)^3}{3k_j{\rm Re}\,T_j^2}\beta_j \gg \beta_j {\rm Re}\,T_j,
\end{align}  
for $j=2,3,4,5$. Thus the tree-level part of 
the gauge kinetic function is always bigger than the one-loop 
corrections of one under the condition that 
${\rm Re}\,T_1 >{\rm Re}\,T_j$ ($j\neq 1$) 
as shown in Eq.~(\ref{eq:gaugekin}), that is, the 
perturbative expansion is valid. 
This property is an important feature of the ``Swiss-Cheese" 
Calabi-Yau manifold. 

From the scalar potential given by using the formula of $4$D $N=1$ 
supergravity, 
\begin{align}
V=e^K\left( K^{I\bar{J}} D_I WD_{\bar{J}} \bar{W} -3|W|^2\right),
\label{eq:scalarpo}
\end{align} 
we get the supersymmetric AdS minimum at the minimum 
given by Eq.~(\ref{eq:m1cd1}),
\begin{align}
\langle V\rangle =-3e^K|W|^2.
\end{align} 
There are several approaches to uplift such AdS vacuum by the 
F-terms with dynamical SUSY breaking 
sector~\cite{Dudas:2006gr,Abe:2006xp,Kallosh:2006dv,Abe:2007yb} or D-terms with anti-heterotic five branes~\cite{Buchbinder:2004im}, 
etc.. 
Here we assume that the SUSY 
is broken by the dynamical SUSY breaking sector 
whose K\"ahler potential and superpotential are given by
\begin{align}
&\Delta K=|X|^2 -\frac{|X|^4}{\Lambda^2},\nonumber\\
&\Delta W=\mu X,
\label{eq:m1KW1loop}
\end{align} 
where X is gauge singlet chiral superfield under the non-abelian 
groups in the visible sector $G_{\rm vis}$ and anomalous $U(1)^{m}$ symmetries, $m=1,2,3,4$, 
$\Lambda$ is the dynamical SUSY breaking scale and we omit the 
moduli dependence of $X$, because they do not affect the following 
moduli stabilization. Then the Minkowski minimum is realized by choosing the parameter $\mu$ as 
\begin{align}
\langle V\rangle +\Delta V\simeq e^{\langle K\rangle}\left( 
-3|\langle W\rangle|^2 +K^{X\bar{X}}|\mu|^2
\right)=0 \Leftrightarrow |\mu|^2 =3|\langle W\rangle|^2.
\end{align} 

Finally, we consider the contribution of the omitted term 
$Ae^{-\frac{8\pi^2}{a}\,\Phi}$ in the superpotential~(\ref{eq:m1KW}) 
which is ignored on the previous analysis. 
Since we assume that such omitted term is much smaller 
than the other terms in the superpotential~(\ref{eq:m1KW}) at the minimum, 
the moduli ${\rm Re}\,\Phi$, $T_1$, ${\rm Re}\,T_2$, ${\rm Re}\,T_3$, ${\rm Re}\,T_4$ and 
${\rm Re}\,T_5$ 
are stabilized at the minimum close to the values given by 
Eq.~(\ref{eq:m1cd1}) and are decoupled from the 
inflaton dynamics if their masses are heavier 
than the inflation scale. 
The mass scales of these moduli are determined by 
the constant term, the world-sheet instanton 
effect of the superpotential~(\ref{eq:m1KW}) 
and the D-term contribution~(\ref{eq:m1K}) 
which is heavier than the inflation scale as shown later. 
(The mass matrices of them 
are summarized in the Appendix~\ref{app:mass}.). 
As mentioned before, the imaginary parts of the moduli 
except for the inflaton $Y^4$ 
are absorbed by the four $U(1)$ gauge bosons 
whose mass scale is of order the string scale $M_s$. 

\subsubsection{Inflaton potential and its dynamics}
Now we are ready to write down the inflation potential. 
As discussed in the previous section~\ref{subsubsec:naturalinf1}, 
after integrating out these heavy 
moduli and substituting the field values given by Eq.~(\ref{eq:m1cd1}), 
we get the effective scalar potential 
for the light moduli $Y^4$ which is the linear combination of 
${\rm Im}\,T_4$ and ${\rm Im}\,T_5$ given by Eq.~(\ref{eq:infdef}), 
\begin{align}
V_{\rm eff} 
\simeq \Lambda^4 (1-{\rm cos}\,(\beta\,\hat{Y}^4)),
\label{eq:ninf_po}
\end{align} 
in the limit of $Ae^{-\frac{8\pi^2}{a}\,\langle {\rm Re}\Phi\rangle}
\ll W_0, Be^{-\mu_1 \langle T_1\rangle}$, 
where the energy scale of the scalar potential $\Lambda^4$ 
and the axion decay constant $\beta$ are defined as 
\begin{align}
\Lambda^4 &\equiv 6\,e^{K}e^{-\frac{8\pi^2}{a}\,{\rm Re}\Phi}
A(W_0+Be^{-\mu_1 T_1}),
\nonumber\\
\beta &\equiv \frac{8\pi^2}{a\,N^4{\hat N}^4}\,
\left(\frac{\beta_5}{q_{T_4}^4\sqrt{K_{T_4\bar{T}_4}}}
-\frac{\beta_4}{q_{T_5}^4\sqrt{K_{T_5\bar{T}_5}}}\right),
\label{eq:decct}
\end{align} 
and 
\begin{align}
{\hat Y^4}\simeq 
\frac{1}{N^4}\sqrt{2\left(\frac{K_{T_4\bar{T}_4}}{(q_{T_5}^4)^2K_{T_5\bar{T}_5}}
+\frac{K_{T_5\bar{T}_5}}{(q_{T_4}^4)^2K_{T_4\bar{T}_4}}\right)}\,Y^4 
\equiv \hat{N}^4\,Y^4
\end{align} 
is the canonically normalized axion field. 
Here we employed the following redefinitions of the 
moduli,
\begin{align}
&{\rm Im}\,T_4=
\frac{1}{N^4}\left( 
\frac{X^4}{q_{T_4}^4\sqrt{K_{T_4\bar{T}_4}}}
-\frac{Y^4}{q_{T_5}^4\sqrt{K_{T_5\bar{T}_5}}}\right),
\nonumber\\
&{\rm Im}\,T_5=
\frac{1}{N^4}\left( 
\frac{X^4}{q_{T_5}^4\sqrt{K_{T_5\bar{T}_5}}}
-\frac{Y^4}{q_{T_4}^4\sqrt{K_{T_4\bar{T}_4}}}\right),
\end{align} 
and $U(1)^4$ gauge invariance of the superpotential~(\ref{eq:m1KW}), 
\begin{align}
q_{T_4}^4\,\beta_4+ q_{T_5}^4\,\beta_5 =0.
\end{align} 

When we identify the axion ${\hat Y}^4$ as the inflaton, 
the effective scalar potential~(\ref{eq:ninf_po}) is considered as the inflation potential for the single-field ${\hat Y}^4$, 
since the mass of the other moduli 
are much heavier than the inflaton. Thus we can realize the 
scalar potential of the type of natural inflation. 
The power spectrum of the scalar density perturbation is 
explained by choosing the parameter,
$\Lambda^4 \sim {\cal O}(10^{-9})$ in the $M_{\rm Pl}$ unit, and 
the spectral index of the scalar density perturbation and 
the tensor-to-scalar ratio is also consistent with the 
cosmological observations reported by WMAP, Planck and/or 
BICEP2 collaborations. This is because we can realize the 
trans-Planckian axion decay constant $\beta$ originating 
from the one-loop corrections to the gauge kinetic function 
as shown in Eq.~(\ref{eq:decct}). 

Next, we estimate the cosmological observables constrained 
by the observations. We choose the dilaton K\"ahler potential 
as the type of $K^0+K^{\rm np}$~\footnote{The stabilization of 
moduli are discussed in Appendix~\ref{app:mass}.} and the following input parameters 
in the K\"ahler potential given by Eqs.~(\ref{eq:m1K}) 
and~(\ref{eq:dilaton}) as 
\begin{align} 
&k_1=k_2=k_3=k_4=k_5=\frac{1}{8},\nonumber\\
&d=7,\,b=1,\,p=2,\nonumber\\
&\beta_2\simeq\beta_3\simeq\beta_4\simeq\beta_5\simeq 0.01,
\label{eq:inputK}
\end{align}  
and in the superpotential given by Eqs.~(\ref{eq:m1W}) 
and~(\ref{eq:m1KW1loop}) as,
\begin{align} 
A\,=\frac{1}{300},\,\,\,a=30,\,\,\,B=-\frac{1}{2},
\,\,\,\mu_1= 2\pi,\,\,\, W_0=6\times 
10^{-4},\,\,\,\mu\simeq 1\times 10^{-3},
\label{eq:inputW}
\end{align}  
in the unit $M_{\rm Pl}=1$ and 
the $U(1)$ charges of the moduli are of ${\cal O}(1)$. 
From these input parameters, 
we get the field values of the moduli at the minimum,
\begin{align} 
T_1\,\simeq 1.3,\,\,\,T_2\simeq T_3\simeq T_4\simeq T_5\simeq 0.06,
\,\,\,
S\simeq \Phi\simeq 2,
\end{align}  
which yield the gauge coupling unification of the 
grand unified theory (GUT) at the Kaluza-Klein (KK) scale, 
\begin{align} 
M_{KK}\simeq \frac{M_s}{{\cal V}^{1/6}} \simeq 1.2\times 10^{17}
\,{\rm GeV},
\end{align}  
with 
\begin{align} 
M_s=\frac{M_{\rm Pl}}{\sqrt{4\pi \alpha^{-1}}} \simeq 1.4\times 10^{17}
\,{\rm GeV},
\end{align}  
where $\alpha^{-1}\simeq 24$ is the gauge coupling 
of visible gauge group $G_{\rm vis}$ at the string 
scale. 

By employing the input parameters given by 
Eqs.~(\ref{eq:inputK}) and~(\ref{eq:inputW}), 
the energy scale of the scalar potential,
\begin{align} 
\Lambda^4 \simeq 3.22 \times 10^{-9}, 
\end{align}  
and the axion decay constant,
\begin{align} 
\beta^{-1} \simeq 7.8,
\end{align}  
in the unit $M_{\rm Pl}=1$, are obtained which leads 
to the desired trans-Planckian axion decay constant. 

To estimate the cosmological observables, we define 
the slow-roll parameters,
\begin{align} 
\epsilon &\equiv \frac{M_{\rm Pl}^2}{2}
\left(\frac{\partial_{\hat{Y}^4}V_{\rm eff}}{V_{\rm eff}}\right)^2,
\nonumber\\
\eta &\equiv M_{\rm Pl}^2 
\frac{\partial_{\hat{Y}^4}^2 V_{\rm eff}}{V_{\rm eff}},
\nonumber\\
\xi^2 &\equiv M_{\rm Pl}^4 
\frac{\partial_{\hat{Y}^4} V_{\rm eff} 
\partial_{\hat{Y}^4}^3 V_{\rm eff}}{V_{\rm eff}^2},
\end{align}  
and then the e-folding number from the time $t_\ast$ 
to the inflation end $t_{\rm end}$ is estimated as 
\begin{align} 
N_e=-\int_{t_{\rm end}}^{t_\ast} dt H(t) \simeq 
\frac{1}{M_{\rm Pl}}\int_{\hat{Y}^4_\ast}^{\hat{Y}^4_{\rm end}}
\frac{d\hat{Y}^4}{\sqrt{2\epsilon}},  
\end{align}  
where the Hubble parameter $H(t)$ is defined as 
$H(t)=\frac{\dot{a}(t)}{a(t)}$, $a(t)$ is the 
scale factor of the $4$D spacetime. 
$\hat{Y}^4_\ast$ and $\hat{Y}^4_{\rm end}$ are the 
field values of the inflaton $\hat{Y}^4$ at the time 
$t_\ast$ and $t_{\rm end}$, respectively.
\footnote{The end of inflation is estimated when the 
slow-roll condition is violated as 
${\rm max}\{ |\epsilon|, |\eta|\}=1$.}
The observables such as the power spectrum 
of the scalar density perturbation~$P_{\zeta}$, 
the spectral index of it~$n_s$, the tensor-to-scalar ratio~$r$ are written in terms of the 
slow-roll parameters as  
\begin{align} 
P_\zeta &=\frac{1}{24\pi^2}
\frac{V_{\rm eff}}{\epsilon M_{\rm Pl}^4},
\nonumber\\
n_s&= 1-6\epsilon +2\eta,
\nonumber\\
r&=16\epsilon.\,\,\,
\end{align}  
At the field value $\hat{Y}^4_\ast \simeq 13M_{\rm Pl}$, 
we find the numerical values of observables and the 
e-folding number as 
\begin{align} 
P_\zeta \simeq 2.2\times 10^{-9},\,\,\,
n_s\simeq 0.956,\,\,\,
r\simeq 0.11,\,\,\,
N_e\simeq 48,
\end{align}  
which are consistent with the WMAP, Planck 
data~\cite{Ade:2013uln},
\begin{align} 
P_\zeta = 2.196^{+0.051}_{-0.060}\times 10^{-9},\,\,\,
n_s= 0.9583\pm 0.0080,
\end{align}  
at the pivot scale $k_{\ast}=0.05 {\rm Mpc}^{-1}$ 
and BICEP2 data~\cite{Ade:2014xna},
\begin{align} 
r=0.16^{+0.06}_{-0.05},
\end{align}  
after considering the foreground dust. 
Now we choose the hidden gauge group 
as $E_8$ which leads to the dual Coxeter number $a=30$ and 
$\beta_3\simeq \beta_4\simeq \beta_5\simeq 0.01$. 

Note that we can realize smaller tensor-to-scalar ratio 
which is more consistent with the WMAP and Planck data, since 
the size of the axion decay constant 
$\beta$ depends on the dual Coxeter number of the hidden 
gauge group~$a$ in Eq.~(\ref{eq:decct}) and the size of one-loop 
correction to the gauge kinetic function of the hidden gauge 
group in Eq.~(\ref{eq:oneloop}).

\subsection{Model $2$ (Double gaugino condensations)}
\label{subsec:naturalinf2}
In this section, we propose the natural inflation based on 
the other type of the K\"ahelr potential and superpotential. 
The main difference between the model $1$ in the previous 
Sec.~\ref{subsec:naturalinf1} and the model $2$ in this section 
is the stabilization mechanism of dilaton. In the model $1$, the 
dilaton is stabilized at the finite value by the non-perturbative 
corrections to its K\"ahler potential given by 
Eq.~(\ref{eq:dilatonK}). 
However, in the model $2$, the dilaton is stabilized 
by using one of the gaugino condensation terms which will be 
mentioned later. 
In the same way as the model $1$, the trans-Planckian 
axion decay constant is realized from the one-loop correction 
to the gauge kinetic function of the hidden gauge group.

\subsubsection{Setup}
We consider the CY manifold expressed by the following 
K\"ahler potential with one $U(1)$ anomalous symmetry,  
\begin{align}
{\cal K}=-&
\ln \left( S+\bar{S} \right)
-\ln 
\left( k_b(T_b+\bar{T}_b)^3 -k_s\left( T_s +\bar{T}_s -\frac{{\cal Q}_s}{2\pi} V_s \right)^3  
-k_s^\prime\left(T_s^\prime +\bar{T}_s^\prime -\frac{{\cal Q}_s^\prime}{2\pi} V_s \right)^3 
\right) ,
\label{eq:m2K}
\end{align} 
in the unit $M_{\rm Pl}=1$, where $h^{1,1}=3$, 
$k_b$, $k_s$, $k_s^\prime$ are positive constants 
determined by the triple intersection number 
$d_{t_bt_bt_b}$, $d_{t_st_st_s}$, 
$d_{t_s^\prime t_s^\prime t_s^\prime}$ 
and $V_s$ is an anomalous $U(1)_s$ vector multiplet 
under which only two moduli 
$T_s$ and $T_s^\prime$ have $U(1)_s$ charge. 
$U(1)_s$ vector multiplet absorbs the linear 
combination of K\"ahler axions, while the other massless axion 
is identified as the inflaton. We further assume 
that the dilaton K\"ahler potential is approximated by its tree-level 
K\"ahler potential. 

Next, we consider the following $U(1)_s$ invariant superpotential,
\begin{align}
W=&w_0+A_2\,e^{-\frac{8\pi^2}{a_2}(S-\beta_s^{(1)} T_s-\beta_s^{\prime (1)} T_s^\prime)} 
+B_2\,e^{-\frac{8\pi^2}{b_2}(S-\beta_s^{(2)} T_s-\beta_s^{\prime (2)} T_s^\prime)} 
+C_2\,e^{-\mu_b T_b}, 
\label{eq:m2W}
\end{align} 
where $w_0$ is the NS flux induced constant term which stabilizes 
the $h^{1,2}$ complex structure moduli of the CY manifold, 
the second and third term of the right handed side (r.h.s.) show the gaugino condensations on two hidden sectors, the fourth term of the (r.h.s.) shows the world-sheet instanton effect on the two-cycle $T_b$.  
Let us discuss the moduli stabilization and the inflaton potential. 

First, $U(1)_s$ vector multiplet becomes massive whose mass scale is 
of order the string scale due to the $U(1)_s$ magnetic fluxes as shown 
in Eq.~(\ref{eq:mass}), and then $U(1)_s$ gauge boson absorbs the linear 
combination of ${\rm Im}~T_s$ and ${\rm Im}~T_s^\prime$ as,
\begin{align}
X_s=\frac{1}{N_s}\left( \frac{{\rm Im}\,T_s}{q_s\sqrt{K_{T_s\bar{T}_s}}} 
+\frac{{\rm Im}\,T_s^\prime}{q_s^\prime\sqrt{K_{T_s^\prime\bar{T}_s^\prime}}}\right), 
\end{align} 
where $N_s=\sqrt{(1/q_s\sqrt{K_{T_s\bar{T}_s}})^2
+(1/q_s^\prime\sqrt{K_{T_s^\prime\bar{T}_s^\prime}})^2}$ with 
$q_s={\cal Q}_s /2\pi$ and $q_s^\prime={\cal Q}_s^\prime /2\pi$ 
and two K\"ahler moduli are canonically normalized under 
the condition that their K\"ahler mixing are neglected, because 
their stabilization is also the same as the previous model $1$ in 
Sec.~\ref{subsec:naturalinf1}. 
Its orthogonal direction (which is identified as the inflaton later), 
\begin{align}
Y_s=\frac{1}{N_s}
\left( -\frac{{\rm Im}\,T_s}{q_s^\prime\sqrt{K_{T_s^\prime\bar{T}_s^\prime}}}
+\frac{{\rm Im}\,T_s^\prime}{q_s\sqrt{K_{T_s\bar{T}_s}}}\right), 
\end{align} 
remains massless. The $U(1)_s$ charges of the moduli are related as 
\begin{align}
&q_s\,\beta_s^{(1)}+q_s^\prime\,\beta_s^{^\prime (1)} =0,
\nonumber\\
&q_s\,\beta_s^{(2)}+q_s^\prime\,\beta_s^{^\prime (2)} =0,
\end{align} 
due to the $U(1)_s$ gauge invariance of the 
superpotential~(\ref{eq:m2W}). 

Second, we redefine a linear combination of dilaton and K\"ahler 
moduli as,
\begin{align}
\Phi =S-\beta_s^{(1)} T_s-\beta_s^{\prime (1)} T_s^\prime,
\end{align} 
and then the K\"ahler potential and superpotential are rewritten by 
\begin{align}
{\cal K}=-&
\ln \left( \Phi +\bar{\Phi} +\beta_s^{(1)} (T_s+\bar{T}_s) 
+\beta_s^{\prime (1)} (T_s^\prime +\bar{T}_s^\prime) \right) \nonumber\\ 
-&\ln \left( k_b(T_b+\bar{T}_b)^3 -k_s\left( T_s +\bar{T}_s -\frac{{\cal Q}_s}{2\pi} V_s \right)^3  
-k_s^\prime\left(T_s^\prime +\bar{T}_s^\prime -\frac{{\cal Q}_s^\prime}{2\pi} V_s \right)^3 
\right),\nonumber\\
W=&w_0+A_2\,e^{-\frac{8\pi^2}{a_2}\,\Phi} 
+B_2\,e^{-\frac{8\pi^2}{b_2}(\Phi +(\beta_s^{(1)} -\beta_s^{(2)}) T_s 
+(\beta_s^{\prime (1)} -\beta_s^{\prime (2)}) T_s^\prime)} +C_2\,e^{-\mu_b T_b}, 
\label{eq:KW2}
\end{align} 
and we assume that first and second and fourth terms of the (r.h.s.) 
in Eq.~(\ref{eq:KW2}) are much larger than the third term of the (r.h.s.) 
in Eq.~(\ref{eq:KW2}) at least at the minimum, that is, 
$w_0$, $A_2\,e^{-\frac{8\pi^2}{a_2}\,\Phi}$, $C_2\,e^{-\mu_b T_b}$ 
$\gg B_2\,e^{-\frac{8\pi^2}{b_2}(\Phi +(\beta_s^{(1)} -\beta_s^{(2)}) T_s 
+(\beta_s^{\prime (1)} -\beta_s^{\prime (2)}) T_s^\prime)}$. 
Such hierarchies between two gaugino condensation terms are 
realized by the differences between the rank of the two hidden 
gauge groups whose gauginos condensate. 
Therefore, at the moment, we ignore the term 
$B_2\,e^{-\frac{8\pi^2}{b_2}(\Phi +(\beta_s^{(1)} -\beta_s^{(2)}) T_s 
+(\beta_s^{\prime (1)} -\beta_s^{\prime (2)}) T_s^\prime)}$ 
in the superpotential~(\ref{eq:KW2}). 

Third, we stabilize the moduli $\Phi$, $T_b$, ${\rm Re}\,T_s$ 
and ${\rm Re}\,T_s^\prime$ by imposing the supersymmetric conditions,
\begin{align}
&D_{\Phi}W=0,\nonumber\\
&D_{T_b}W=0,\nonumber\\
&K_{T_s}=K_{T_s^\prime}=0,
\label{eq:m2cd1}
\end{align} 
which leads to the vanishing $D$-terms induced from the K\"ahler 
potential~(\ref{eq:KW2}). 
From the scalar potential in the framework of $4$D $N=1$ 
supergravity given by Eq.~(\ref{eq:scalarpo}), 
we get the supersymmetric AdS minimum at the minimum 
given by Eq.~(\ref{eq:m2cd1}), 
\begin{align}
\langle V\rangle =-3e^K|W|^2.
\end{align} 
In the same way as the previous model $1$ in the Sec.~\ref{subsec:naturalinf1}, 
here we assume that the dynamical 
SUSY breaking sector uplift this AdS minimum. 
Their K\"ahler potential and superpotential are given by
\begin{align}
&\Delta K=|X|^2 -\frac{|X|^4}{\Lambda^2},\nonumber\\
&\Delta W=\mu X,
\label{eq:m2KW1loop}
\end{align} 
where X is the gauge singlet chiral superfield under the non-abelian 
gauge groups $G_{\rm vis}$ and anomalous $U(1)_s$ symmetry, 
$\Lambda$ is 
the dynamical SUSY breaking scale and we omit the moduli 
dependence of $X$, because they do not affect the following 
moduli stabilization. 
The Minkowski minimum is realized by choosing the parameter 
$\mu$ as 
\begin{align}
\langle V\rangle +\Delta V\simeq e^K\left( -3|W|^2 +K^{X\bar{X}}|\mu|^2
\right)=0,\,\,\,
\Leftrightarrow |\mu|^2= 3|\langle W\rangle|^2.
\end{align} 

Finally, we consider the term 
$B_2\,e^{-\frac{8\pi^2}{b_2}(\Phi +(\beta_s^{(1)} -\beta_s^{(2)}) T_s 
+(\beta_s^{\prime (1)} -\beta_s^{\prime (2)}) T_s^\prime)}$ in the 
superpotential~(\ref{eq:KW2}). Since we assume that 
such term is much smaller than the other terms 
in the superpotential~(\ref{eq:KW2}), the moduli 
$\Phi$, $T_b$, ${\rm Re}T_s$, ${\rm Re}T_s^\prime$ 
are stabilized at the values close to the minimum given by 
Eq.~(\ref{eq:m2cd1}) and they become massive 
due to the constant term of the superpotential 
for $\Phi$, ${\rm Re}T_s$ and ${\rm Re}T_s^\prime$ 
and the world-sheet instanton effect for $T_b$. 
As ${\rm Re}T_s$ and ${\rm Re}T_s^\prime$, they also obtain 
the D-term contributions from the K\"ahler 
potential~(\ref{eq:KW2}). 

\subsubsection{Inflaton potential}
Let us discuss the inflaton potential. 
After integrating out these heavy moduli and 
substituting the field values 
given by Eq.~(\ref{eq:m2cd1}), we get the effective scalar 
potential for the light modulus $Y_s$ which is the linear 
combination of ${\rm Im}\,T_s$ and ${\rm Im}\,T_s^\prime$,
\begin{align}
V_{\rm eff} 
\simeq \Lambda_s^4 (1-{\rm cos}\,(\beta_s\,\hat{Y}_s)),
\end{align} 
in the limit of $B_2e^{-\frac{8\pi^2}{b_2}\,\langle{\rm Re}\Phi\rangle} 
\ll w_0, A_2e^{-\frac{8\pi^2}{a_2}\,\langle{\rm Re}\Phi\rangle}, 
C_2e^{-\mu_b \langle T_b\rangle}$, 
where 
\begin{align}
\Lambda_s^4\equiv 6e^{K}e^{-\frac{8\pi^2}{b_2}\,{\rm Re}\Phi}B_2
(w_0+A_2\,e^{-\frac{8\pi^2}{a_2}\,\Phi}+C_2\,e^{-\mu_b T_b}),
\end{align}
and the axion decay constant $\beta_s$ is defined 
by 
\begin{align}
\beta_s \equiv \frac{8\pi^2}{b_2N_s{\hat N}_s}\,
\left( -\frac{\beta_s^{(1)}-\beta_s^{(2)}}
{q_{s}^\prime \sqrt{K_{T_s^\prime\bar{T}_s^\prime}}}+
\frac{\beta_s^{\prime (1)}-\beta_s^{\prime (2)}}
{q_s\sqrt{K_{T_s\bar{T}_s}}}\right).
\end{align} 
${\hat Y}_s$ is the canonically normalized axion field, 
\begin{align}
{\hat Y}_s\simeq 
\frac{1}{N_s}\sqrt{2\left (\frac{K_{T_s\bar{T}_s}}{(q_{s}^\prime)^2 
K_{T_s^\prime\bar{T}_s^\prime}} 
+\frac{K_{T_s^\prime\bar{T}_s^\prime}}{(q_{s})^2 
K_{T_s\bar{T}_s}}\right)}\,Y_s 
\equiv \hat{N}_s\,Y_s.
\end{align} 
Here we employed the following redefinitions of the 
moduli,
\begin{align}
&{\rm Im}\,T_s=
\frac{1}{N_s}\left( \frac{X_s}{q_{s}\sqrt{K_{T_s\bar{T}_s}}}
-\frac{Y_s}{q_{s}^\prime\sqrt{K_{T_s^\prime\bar{T}_s^\prime}}}
\right),
\nonumber\\
&{\rm Im}\,T_s^\prime=
\frac{1}{N_s}\left( 
\frac{X_s}{q_{s}^\prime\sqrt{K_{T_s^\prime\bar{T}_s^\prime}}} 
+\frac{Y_s}{q_{s}\sqrt{K_{T_s\bar{T}_s}}}\right),
\end{align} 
and $U(1)_s$ gauge invariance of the superpotential~(\ref{eq:m2W}), 
\begin{align}
&q_s\,\beta_s^{(1)}+q_{s}^\prime\,\beta_s^{^\prime (1)} =0,
\nonumber\\
&q_s\,\beta_s^{(2)}+q_{s}^\prime\,\beta_s^{^\prime (2)} =0.
\end{align} 

Thus when we consider the axion ${\hat Y}_s$ as the inflaton, 
the effective scalar potential is the type of natural inflation. 
The power spectrum of the scalar density perturbation is 
explained by choosing the parameter,
\begin{align} 
\Lambda_s^4 \sim {\cal O}(10^{-9})
\end{align}  
in the $M_{\rm Pl}$ unit, and 
the spectral index of the scalar density perturbation and 
the tensor-to-scalar ratio are also consistent with the 
cosmological observations reported by WMAP, Planck and 
BICEP2 collaborations. This is because we can realize the 
trans-Planckian axion decay constant $\beta_s$ originating 
from the one-loop corrections to the gauge kinetic function. 

However, $E_8\times E_8$ or $SO(32)$ heterotic string 
theories have the rank $16$ gauge groups which have to 
incorporate the rank $4$ SM gauge groups. Then the energy 
scales which two gaugino condense are constrained since 
the total rank of their gauge groups are taken up to $12$ included 
in $E_8\times E_8$ or $SO(32)$. Thus we would need tuning 
some  parameters to realize the correct inflation scale.

\section{Conclusion}
\label{sec:con}
In this paper, we proposed two natural inflation scenarios 
based on the weakly coupled $E_8\times E_8$ or $SO(32)$ 
heterotic string theory on the ``Swiss-cheese" Calabi-Yau 
manifold with multiple $U(1)$ magnetic fluxes. 
The natural inflation is consistent with the WMAP, Planck 
and/or BICEP2 data, only if the size of axion decay constant 
becomes the trans-Planckian. 
However, such trans-Planckian axion decay 
constant is problematic from the theoretical point of view, 
especially on the supergravity models or the string theory. 
So far, there are known scenarios to get the 
trans-Planckian axion decay constant from the 
sub-Planckian axion decay constants~\cite{Kim:2004rp}. 

We identified the inflaton as one of the linear combination of 
K\"ahler axions associated with the two-cycles of the 
Calabi-Yau manifold. When the gauginos of the hidden 
gauge group are condensed, the gaugino condensation terms 
are generated on the superpotential in the framework 
of $4$D ${\cal N}=1$ supergravity. 
In this case, we can realize the trans-Planckian axion 
decay constant originating from the one-loop 
corrections to the gauge kinetic function of the hidden 
gauge group derived from one-loop Green-Schwarz 
term~\cite{Ibanez:1986xy} which is the feature of 
the weakly coupled heterotic string theory. 
On the other hand, in type II superstring theory 
such as the intersecting D-models or magnetized 
D-branes, the gauge kinetic function has the 
${\cal O}(1)$ moduli mixing induced from the winding 
number of D-brane, magnetic fluxes or 
instanton effects.

To realize the single-field inflaton potential, we have to 
stabilize the dilaton and the other K\"ahler moduli. 
At the same time, their masses should be heavier than 
the inflation scale, 
otherwise these moduli would be oscillated during and 
after the inflation which may lead to the sizable 
isocurvature perturbations and cosmological moduli 
problem. Therefore, we 
considered two stabilization scenarios 
categorized as the model $1$ and $2$ based on the 
$E_8\times E_8$ or $SO(32)$ 
heterotic string theory with multiple 
$U(1)$ magnetic fluxes.

In the case of model $1$ discussed in the 
Sec.~\ref{subsec:naturalinf1}, 
the dilaton is stabilized at the finite value by the 
contributions from its non-perturbative 
corrections to the K\"ahler potential. The volume 
moduli is also stabilized by the world-sheet instanton 
effect which leads to the stabilization of the other 
real parts of K\"ahler moduli by using the nature of 
``Swiss-cheese" Calabi-Yau manifold. 
By employing the multiple $U(1)$ magnetic fluxes, 
the imaginary parts of the moduli except for the inflaton 
are absorbed by the corresponding anomalous $U(1)$ 
gauge bosons and then they become massive which 
is of order the string scale. 
Thus we can realize the single-field axion potential 
with trans-Planckian axion decay constant 
determined by the one-loop corrections to the 
gauge kinetic function of the hidden gauge group. 

The essential difference between the model $1$ and $2$ 
is the stabilization mechanism of dilaton. In model $2$ 
discussed in the Sec.~\ref{subsec:naturalinf2}, 
the dilaton is stabilized by one of the gaugino condensation 
terms and we get the effective scalar potential for a linear 
combination of the K\"ahler axions. 
These two proposed inflation scenarios are consistent 
with the WMAP, Planck and/or BICEP2 data, although 
we need to tune the parameters in the model $2$. 

We can also realize smaller tensor-to-scalar ratio 
which is more consistent with the WMAP and Planck data, since 
the size of axion decay constant depends on the dual 
Coxeter number and the size of one-loop 
correction to the gauge kinetic function of the hidden gauge 
group in the heterotic string theory.

\subsection*{Acknowledgement}
H.A. was supported in
part by the Grant-in-Aid for Scientific Research No. 25800158 from the
Ministry of Education,
Culture, Sports, Science and Technology (MEXT) in Japan. T.K. was
supported in part by
the Grant-in-Aid for Scientific Research No. 25400252 from the MEXT in
Japan.
H.~O. was supported in part by a Grant-in-Aid for JSPS Fellows 
No. 26-7296 and a Grant for 
Excellent Graduate Schools from the MEXT in Japan.

\appendix
\section{Mass matrices in model $1$}
\label{app:mass}
In this appendix, we show the mass-squared matrices 
of the scalar potential given by the K\"ahler potential~(\ref{eq:m1KW}) 
and superpotential~(\ref{eq:m1KW}) in the model $1$. 
As we have seen in Sec. \ref{subsec:naturalinf1}, the 
moduli are stabilized at the value given by the supersymmetric 
conditions, $K_I=0$ with 
$I=\Phi,T^2,T^3,T^4$ and they will become massive due 
to the constant superpotential and D-term contributions 
as shown later. 
For completeness, we assume the ansatz of the dilaton 
K\"ahler potential such as 
$K=K^0+K^{\rm np}$ in Eq.~(\ref{eq:dilatonK}). 

First, we canonically normalize the moduli to estimate 
their masses. 
In the case of K\"ahler potential~(\ref{eq:m1KW}) whose dilaton 
K\"ahler potential is replaced with 
$K=K^0+K^{\rm np}$ 
in Eq.~(\ref{eq:dilatonK}), the non-vanishing 
K\"ahler mixing of the dilaton and 
K\"ahler moduli are expanded in the limit of 
${\rm Re}\,S \gg \beta_j {\rm Re}\,T_j$ and $T_1 \gg T_j$, 
$j=2,3,4,5$,   
\begin{align}
K_{\Phi \bar{\Phi}}&\simeq 
-\frac{b}{16}\frac{2}{(\Phi+\bar{\Phi})^{3/2}}K^{\rm np} 
+\frac{1}{2}\left( p-b\left(\frac{\Phi+\bar{\Phi}}{2}\right)^{1/2}\right) 
\frac{1}{(\Phi+\bar{\Phi})^{2}},
\nonumber\\
K_{\Phi \bar{T}_j}&\simeq 
\frac{\beta_j}{(\Phi+\bar{\Phi})^{2}}, 
\nonumber\\
K_{T_1 \bar{T}_1}&\simeq 
\frac{3}{(T_1+\bar{T}_1)^2}, 
\nonumber\\
K_{T_1\bar{T}_j}&\simeq 
\frac{9k_j(T_j+\bar{T}_j)^2}{k_1(T_1+\bar{T}_1)^4},
\nonumber\\
K_{T_j\bar{T}_j}&\simeq 
\frac{6k_j(T_j+\bar{T}_j)}{k_1(T_1+\bar{T}_1)^3},
\nonumber\\
K_{T_i\bar{T}_j}&\simeq 
\frac{9k_ik_j(T_i+\bar{T}_i)^2(T_j+\bar{T}_j)^2}{k_1^2(T_1+\bar{T}_1)^6},
\label{eq:Kmetric}
\end{align}
with $i\neq j$, $i,j=2,3,4,5$. Here we use the following 
stabilization conditions of the moduli,
\begin{align}
K_{\Phi}&\simeq 
-\frac{1}{\Phi+\bar{\Phi}} 
+\frac{1}{2(\Phi+\bar{\Phi})}
\left( p-b\left(\frac{\Phi+\bar{\Phi}}{2}\right)^{1/2}\right) 
K^{\rm np}=0,
\nonumber\\ 
K_{T_j}&\simeq 
\frac{3k_j(T_j+\bar{T}_j)^2}{k_1(T_1+\bar{T}_1)^3} 
-\frac{\beta_j}{\Phi+\bar{\Phi}}=0, 
\label{eq:stcd}
\end{align}
for $j=2,3,4,5$. 
As discussed in Sec.~\ref{subsec:naturalinf1}, 
the perturbative expansion is ensured under 
the above stabilization conditions, 
that is, $S\gg \beta_jT_j$ for $j=2,3,4,5$.  
Since the off-diagonal elements 
are suppressed by the smallness of $\beta_j$ 
and the value of moduli $T_j$, $j=2,3,4,5$ 
at the minimum given by Eq.~(\ref{eq:stcd}), 
the moduli K\"ahler metric are approximated by their 
diagonal form,
\begin{align}
K_{I\bar{J}}\simeq K_{I\bar{J}}\delta_{I\bar{J}},
\end{align}
with $I,J=\Phi,T_1,T_j$ for $j=2,3,4,5$.

Second, we show the mass matrices given by 
the D-term potential 
which is obtained from the K\"ahler 
potential~(\ref{eq:m1KW}) whose dilaton 
K\"ahler potential is replaced with 
$K=K^0+K^{\rm np}$ 
in Eq.~(\ref{eq:dilatonK}),
\begin{align}
V_D=&\frac{1}{2f_{U(1)^1}}(q_S^1K_S +q_{T_2}^1K_{T_2}
+q_{T_3}^1K_{T_3})^2
+\frac{1}{2f_{U(1)^2}}(q_S^2K_S +q_{T_2}^2K_{T_3} 
+q_{T_3}^2K_{T_3})^2 \nonumber\\
&+\frac{1}{2f_{U(1)^3}}(q_S^3K_S+q_{T_2}^3K_{T_2} +q_{T_3}^3K_{T_3})^2
+\frac{1}{2f_{U(1)^4}}(q_{T_4}^4K_{T_4} +q_{T_5}^4K_{T_5})^2,
\end{align}
where the gauge kinetic functions of $U(1)^m$, $m=1,2,3,4$ 
are approximated as $f_{U(1)^m}\simeq {\rm tr}(T^mT^m)S$. 
At the SUSY minimum where $K_I=0$ with 
$I=\Phi,T_2,T_3,T_4,T_5$ and $D_{T_1}W=0$, 
the second 
derivatives of the above D-term potential can be 
expanded in the small parameter $\beta_j$, $j=2,3,4,5$ as 
\begin{align}
(V_D)_{I\bar{J}}=(V_D)_{I\bar{J}}^{0}+(V_D)_{I\bar{J}}^{1}+\cdots
\end{align}
where 
\begin{align}
(V_D)_{\Phi\bar{\Phi}}^{0} =&\sum_{n=1}^3\frac{1}{2f_{U(1)^n}} 
(q_S^nK_{\Phi\bar{\Phi}}+q_{T_2}^nK_{T_2\bar{\Phi}}+ 
q_{T_3}^nK_{T_3\bar{\Phi}})^2,
\nonumber\\
(V_D)_{\Phi\bar{T}_2}^{0} =&\sum_{n=1}^3\frac{1}{2f_{U(1)^n}}
(q_S^nK_{\Phi\bar{\Phi}}+q_{T_2}^nK_{T_2\bar{\Phi}})q_{T_2}^n
K_{T_2\bar{T}_2}, 
\nonumber\\
(V_D)_{\Phi\bar{T}_3}^{0} =&\sum_{n=1}^3\frac{1}{2f_{U(1)^n}}
(q_S^nK_{\Phi\bar{\Phi}}+q_{T_3}^nK_{T_3\bar{\Phi}})q_{T_3}^n
K_{T_3\bar{T}_3}, 
\nonumber\\
(V_D)_{T_2\bar{T}_2}^{0} =&\sum_{n=1}^3\frac{1}{2f_{U(1)^n}} 
(q_{T_2}^n)^2 (K_{T_2\bar{T}_2})^2, 
\nonumber\\
(V_D)_{T_2\bar{T}_3}^{0} =&\sum_{n=1}^3\frac{1}{2f_{U(1)^n}} 
q_{T_2}^n q_{T_3}^n K_{T_2\bar{T}_2} K_{T_3\bar{T}_3}, 
\nonumber\\
(V_D)_{T_3\bar{T}_3}^{0} =&\sum_{n=1}^3\frac{1}{2f_{U(1)^n}} 
(q_{T_3}^n)^2 (K_{T_3\bar{T}_3})^2, 
\nonumber\\
(V_D)_{T_4\bar{T}_4}^{0} =&\frac{1}{2f_{U(1)^4}} 
\left( q_{T_4}^4K_{T_4\bar{T}_4}  +q_{T_5}^4K_{T_4\bar{T}_5}\right)^2, 
\nonumber\\
(V_D)_{T_4\bar{T}_5}^{0} =&\frac{1}{2f_{U(1)^4}}
\left( q_{T_4}^4K_{T_4\bar{T}_4}  +q_{T_5}^4K_{T_4\bar{T}_5}\right)
\left( q_{T_4}^4K_{T_4\bar{T}_5}  +q_{T_5}^4K_{T_5\bar{T}_5}\right), 
\nonumber\\
(V_D)_{T_5\bar{T}_5}^{0} =&\frac{1}{2f_{U(1)^4}}
\left( q_{T_4}^4K_{T_4\bar{T}_5}  +q_{T_5}^4K_{T_5\bar{T}_5}\right)^2,  
\end{align}
and the other elements of the mass matrices are vanishing. 
As can be seen in Eq.~(\ref{eq:Kmetric}), 
$(V_D)_{I\bar{J}}^{0}$ are of order $\beta_j^2$, $j=2,3,4,5$. 
On the other hand, we find that $(V_D)_{I\bar{J}}^{1}$ and 
$(V_D)_{I\bar{J}}^{2}$ are of order $\beta_j^3$ and $\beta_j^4$, 
respectively and they are smaller than the mass terms obtained 
from the F-term contributions in the case of 
input parameters given by Eqs.~(\ref{eq:inputK}) 
and (\ref{eq:inputW}).

The mass matrices given by the F-term potential 
which is obtained from the K\"ahler potential 
and superpotential~(\ref{eq:m1KW}) are shown as 
\begin{align}
(V_F)_{\Phi\bar{\Phi}}\simeq &e^K K^{\Phi\bar{\Phi}}
|K_{\Phi\bar{\Phi}}W|^2,
\nonumber\\
(V_F)_{T_1\bar{T}_1}\simeq &e^K K^{T_1\bar{T}_1}
|W_{T_1}|^2,
\nonumber\\
(V_F)_{T_2\bar{T}_2}\simeq &e^K K^{T_2\bar{T}_2}
|K_{T_1\bar{T}_1}W|^2,
\nonumber\\
(V_F)_{T_3\bar{T}_3}\simeq &e^K K^{T_3\bar{T}_3}
|K_{T_3\bar{T}_3}W|^2,
\nonumber\\
(V_F)_{T_4\bar{T}_4}\simeq &e^K K^{T_4\bar{T}_4}
|K_{T_4\bar{T}_4}W|^2,
\nonumber\\
(V_F)_{T_5\bar{T}_5}\simeq &e^K K^{T_5\bar{T}_5}
|K_{T_5\bar{T}_5}W|^2,
\end{align} 
and other elements of the mass matrices are vanishing 
at the minimum given by Eqs.~(\ref{eq:m1cd1}). 
Here the K\"ahler metric is approximated as 
the diagonal form and we neglect the gaugino condensation 
term in Eq.~(\ref{eq:m1KW}). 

Finally we show the total mass matrices 
given by the D-term and F-term potential 
approximated as
\begin{align}
(V)_{I\bar{J}}\simeq (V_D)_{I\bar{J}}^{0}+(V_F)_{I\bar{J}},
\end{align} 
and we find that this mass-squared matrices are full-rank and 
the eigenvalues of them are positive in the choice 
of the input parameters. Their mass scales of the moduli are 
determined by the string scale and 
SUSY breaking scale~$m_{3/2}=e^{\langle K\rangle/2}\langle W\rangle 
\simeq 5\times 10^{14} {\rm GeV}$ from the D-term and F-term 
contributions, respectively.

\end{document}